\documentclass[aps,prl,twocolumn,groupedaddress,showpacs]{revtex4}
\usepackage{graphicx}

\newcommand{\ef}{E_{\scriptscriptstyle F}}
\newcommand{\ut}{\widetilde{U}}

\newcommand{\leqa}{\stackrel{<}{\scriptstyle \sim}}

\begin{document}

\title{Correlations in Nuclear Masses}

\author{H. Olofsson,$^1$ S. \AA berg,$^1$ O. Bohigas,$^2$ and P. Leboeuf$~^2$}

\affiliation{$^1$Division of Mathematical Physics, LTH, Lund University. P.O.
  Box 118, S-221 00 Lund, Sweden \\ $^2$Laboratoire de Physique Th{\'e}orique
  et Mod{\`e}les Statistiques$^*$, B{\^a}timent 100, Universit{\'e} de
  Paris-Sud, 91405 Orsay Cedex, France}

\date{\today}

\begin{abstract}
It was recently suggested that the error with respect to experimental data in
nuclear mass calculations is due to the presence of chaotic motion. The theory
was tested by analyzing the typical error size. A more sensitive quantity, the
correlations of the mass error between neighboring nuclei, is studied here.
The results provide further support to this physical interpretation.
\end{abstract}


\maketitle

The precision of nuclear mass spectrometry has dramatically improved in recent
years. Thanks to the Penning trap technique, mass measurements with relative
precision as high as $10^{-8}$, or even $10^{-10}$ for stable nuclei, may be
achieved \cite{massacc}. This unprecedented accuracy has important
consequences in different areas, like the determination of fundamental
constants, symmetry violations, metrology, stellar evolution and
nucleosynthesis.

An additional motivation for such precise measurements is to establish
accurate and predictive mass formulae. Global nuclear mass calculations have
been pursued over the years with increasing accuracy \cite{mnms,gtp,dz} (see
Ref.\cite{lpt} for a review on recent experimental and theoretical
developments). Despite the numerous parameters contained in the different
models and the variety of the approaches adopted, two peculiar features emerge
from these calculations. On the one hand, different models yield similar
results for the known masses. A typical accuracy is $5 \times 10^{-4}$ for a
medium--heavy nucleus whose total (binding) energy is of the order of 1000
MeV. On the other hand, the predictions of different mass models strongly
diverge when applied to unknown regions (they may differ by several MeV, i.e.
relative variations of order $5 \times 10^{-3}$).

These two features point towards the possibility of a basic underlying
physical mechanism not appropriately incorporated into the present models.
This mechanism should explain, in particular, the observed differences between
measured and calculated masses. In Ref.\cite{bl} it was shown that the
presence of chaotic layers in the nucleonic motion, whatever its physical
origin may be, leads to a contribution to the nuclear mass whose typical size
$\sigma_{ch}$ is given by
\begin{equation} \label{sigmach}
\sigma_{ch} = \frac{2.8}{A^{1/3}} \ {\rm MeV} \ ,
\end{equation}
where $A$ is the mass number. Equation (\ref{sigmach}), obtained through a
mean--field theory, follows from very general arguments, and is independent of
any detailed information concerning the system. It provides an
order--of--magnitude estimate of the chaotic contribution, and determines the
onset of a new regime. Figure~1 shows, as a function of $A$, the comparison
between this prediction and the typical size of the error of several nuclear
mass formulae using the most recent experimental compilation \cite{awt}. Three
different global mass models are compared. Two of them are based on mean field
theory. The first one is a non-self-consistent macroscopic-microscopic model
\cite{mnms}, the second one is a self-consistent calculation based on
Hartree-Fock-BCS \cite{gtp} while the third one \cite{dz} is a shell--model
based calculation with parameterized monopole and multipole terms. The
agreement with Eq.(\ref{sigmach}) for the two mean--field models is
remarkable. The mass number dependence of the error is also well described for
the third model with, however, a factor of order two between their amplitudes,
a tendency that one might expect for a model that includes substantial
residual interaction effects.

\begin{figure} \label{rmsmassA}
\centerline{\includegraphics[width=7.5cm]{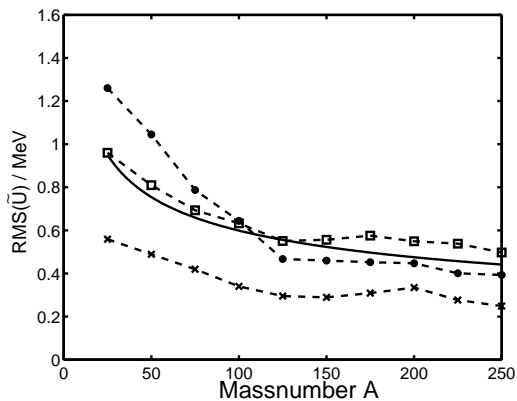}}
\caption{Root mean square of the difference between experimentally measured
and calculated masses. Dots, squares and crosses from the calculations of
Refs.\cite{mnms}, \cite{gtp} and \cite{dz}, respectively. The solid line shows
the chaotic contribution, Eq.(\ref{sigmach}).}
\end{figure}

When the difference between calculated and measured masses is plotted as a
function of the neutron number (or of any other relevant parameter), an
oscillatory curve, whose typical amplitude is shown in Fig.1, is observed.
This curve doesn't look as a random white noise signal but shows structures.
This qualitative remark is consistent with theoretical expectations. Indeed,
it is known from semiclassical mean--field theories \cite{lm,bl} that the
fluctuations of the mass produced by chaotic layers are dominated by the short
classical orbits. The statistical properties of these orbits show system
dependent features. As a consequence, definite non--random fluctuations for
the chaotic component of the mass are expected. Some of the predictions, like
the asymmetry and non--Gaussian nature of the probability distribution of the
oscillations \cite{lm}, have been tested recently \cite{hfv}. Thus, in the
present context chaos should not be assimilated to a random unpredictable
process. Though it may be a difficult task to explicitly compute its
contribution for each nucleus \cite{aberg} because, generically, the detailed
oscillatory structure of this contribution is very sensitive to details of the
Hamiltonian, it does not set an a priori bound to the accuracy of the
theoretical mass calculations \cite{rydberg}. This view differs from that
expressed in Refs.\cite{mw}, where it is stated that the (partially) chaotic
nuclear motion imposes limits {\sl in principle} on the accuracy of the
calculations of binding energies (cf Ref.\cite{aberg}). Moreover, and in
contrast to global microscopic mass models, seemingly random deviations of
typical size $\leqa 100$ keV are observed in algebraic mass relations designed
to locally cancel the interactions between nucleons \cite{bfhv}.

Our aim here is to further develop the analysis of the mass errors in terms of
chaotic motion by computing the autocorrelation function at different neutron
numbers. Compared to the estimate of the typical error size, this is a finer
analysis of the fluctuations, that tests more subtle dynamical information. As
discussed below, by assuming the errors are due to the contribution of chaotic
motion it is possible to obtain definite predictions which, in particular, fix
a typical scale of the correlations.

Fluctuations of the nuclear masses may be written, in a semiclassical
mean--field expansion, as \cite{bl}
\begin{equation} \label{ut}
\widetilde{U} (x) = 2 \hbar^2 \sum_p \sum_{r=1}^{\infty}
\frac{A_{p,r}}{r^2 ~ \tau_p^2} \cos(r S_p/\hbar+\nu_{p,r})  \ .
\end{equation}
The sum is over all the primitive periodic orbits $p$ (and their repetitions
$r$) of a classical underlying effective single--particle Hamiltonian. Each
orbit is characterized by its action $S_p$, stability amplitude $A_{p,r}$,
period $\tau_p=\partial S_p/\partial E$ and Maslov index $\nu_{p,r}$. $x$ is a
parameter on which the effective potential depends on. Though we let it for
the moment unspecified, it will be chosen below to be the number of neutrons.
The orbits entering this expression are all evaluated at the Fermi energy
$\ef$. The latter is related to the mass number through the condition
$\int_0^{\ef} \overline{\rho} (E,x) ~ dE = A$, where $\overline{\rho}$ is the
average single--particle density of states.

When the parameter $x$ is varied, the correlation function of the energy
fluctuations $\widetilde{U}$ is defined as,
\begin{equation}\label{cdef}
C(x) = \langle\widetilde{U}(x_0-x/2) \ \widetilde{U}(x_0+x/2)\rangle_{x_0} \ ,
\end{equation}
where the brackets denote an average over a suitable parameter window, which
is large compared to the typical scales of oscillation of $\ut$ and small on a
classical scale. The correlation is evaluated using Eq.(\ref{ut}). When the
parameter $x$ varies, and for a large number of particles where $S_p \gg
\hbar$, the main contributions to the variations of $\widetilde{U}$ come from
the variations of the action. In a linear approximation, valid when the
variations are large compared to $\hbar$ but small compared to $S_p$, the
action varies as $S(x_0 \pm x/2) = S(x_0) \pm Q_p x/2$, where $Q_p = \partial
S_p /\partial x|_{x_0}$. Then, from (\ref{cdef}) and (\ref{ut}), dropping terms
whose average is zero, we obtain
\begin{eqnarray}  \label{cds}
  C(x) & = & 2\hbar^4\left< \sum_{p,p'}\sum_{r,r'}\frac{A_{p,r}}{r^2\tau_p^2}
  \frac{A_{p',r'}}{r'^2\tau_{p'}^2}\right. \times \\ && \left.\cos\left(
  \frac{rS_p-r'S_{p'}}{\hbar}\right) \cos
  \left(\frac{rQ_p+r'Q_{p'}}{2\hbar}x\right)\right>_{x_0} . \nonumber
\end{eqnarray}
The double sum contains interfering terms between different orbits. However,
as shown in Ref.\cite{lm}, it is not necessary to consider them. The main
contributing orbits are in fact the shortest ones (i.e, those having the
shortest period). Their contribution is well approximated by taking into
account only the diagonal terms $p=p'$, $r=r'$,
\begin{eqnarray} \label{cdiag1}
C(x) & = & 2\hbar^4 \sum_{p,r}\frac{A_{p,r}^2}{r^4\tau_p^4}
cos\left(\frac{rQ_p}{\hbar}x\right) \\ \label{cdiag2}
& = & \frac{\hbar^2}{2\pi^2}\int_0^\infty\frac{d\tau}{\tau^4} \left
<\cos\left(\frac{Qx}{\hbar}\right)\right>_\tau K_{\scriptscriptstyle D} (\tau)
\ .
\end{eqnarray}
In Eq.(\ref{cdiag2}) we have used the semiclassical expression of the spectral
form factor, Eq.(\ref{ffd}) below, to express the autocorrelation in terms of
an integral over time. The average over the cosine function is computed over
the whole set of periodic orbits of period between $\tau$ and $\tau + d \tau$.
Denoting $P_\tau(Q)$ the distribution of the values of $Q$ of all these
orbits, then
\begin{equation} \label{ca}
\left<\cos\left(Qx/\hbar \right)\right>_\tau=\int_{-\infty}^\infty
\cos\left(Qx/\hbar \right)P_\tau(Q)dQ.
\end{equation}

Equations (\ref{cdiag1}) and (\ref{cdiag2}) are valid for both regular and
chaotic motion. To proceed further, we must specify the type of dynamics we
are considering. As in \cite{bl}, we identify the error between experimental
($U_{\rm exp}$) and calculated mass ($U_{\rm calc}$) as originating from the
chaotic contribution ($\widetilde{U}$) of the nuclear motion, $U_{\rm exp} =
U_{\rm calc} + \widetilde{U}$. When the motion is chaotic, the number of
periodic orbits having period between $\tau$ and $\tau + d \tau$ grows
exponentially with $\tau$. There are numerical as well as theoretical
evidences indicating that the distribution function $P_\tau(Q)$ is in this
case Gaussian, of average $\langle Q \rangle_\tau = \eta \tau$ and width
$\langle Q^2 \rangle_\tau = \alpha \tau$ \cite{ls}. Nontrivial dynamical
information is contained in $\alpha$. When $x$ represents variations of the
neutron number, the parameter $\eta$ takes into account the effect of the
increase of the volume of the nucleus as a neutron is added, implying an
increase of the average action (or length) of the orbits. Finally, an explicit
expression for $K_D(\tau)$ is also needed. The semiclassical expression for
$K_D(\tau)$ contains detailed information about the discrete spectrum of
periods $\tau_p$ of the periodic orbits,
\begin{equation} \label{ffd}
K_{\scriptscriptstyle D} (\tau) =
h^2\sum_{p,r}A_{p,r}^2\delta(\tau-r\tau_p) \ .
\end{equation}
Here, we will not take into account this detailed information, but instead
use, as in \cite{bl}, a continuous approximation \cite{lm}
\begin{equation} \label{ffa}
K_{\scriptscriptstyle D} (\tau) \approx \left\{
\begin{array}{lll}
K_{\scriptscriptstyle D} (\tau) &=& 0 \;\;\;\;\;\;\;\;\; \tau < \tau_{min} \\
K_{\scriptscriptstyle D} (\tau) &=& 2 \ \tau \;\;\;\;\;\; \tau \geq \tau_{min} \
, \\ \end{array} \right.
\end{equation}
where $\tau_{min}$ is the period of the shortest periodic orbit of the system.
This schematic approximation incorporates two important features. On the one
hand, it contains a system dependent information, namely the truncation for
times smaller than $\tau_{min}$. On the other hand, it displays the
universality observed in chaotic systems with time reversal invariance for
times $\tau_{min} \ll \tau \ll h \overline{\rho}$, namely the linear growth
characteristic of random matrix theory.

Using the Gaussian form of $P_\tau(Q)$ in Eq.(\ref{ca}) as well as the
approximation (\ref{ffa}) of $K_{\scriptscriptstyle D} (\tau)$, the
correlation function Eq.(\ref{cdiag2}), normalized to one at the origin, takes
the form
\begin{equation} \label{cna}
C (\zeta)=\frac{\zeta^4}{2}\int_{\zeta^2/2}^\infty\frac{dy}{y^3}cos\left(
  \frac{a y}{\zeta}\right) e^{-y/2},
\end{equation}
where the dimensionless parameters $\zeta$ and $a$ are given by
\begin{equation} \label{unf}
\zeta = \frac{\sqrt{2 \ \alpha \ \tau_{min}}}{\hbar} \ x \;\;\;\;\; {\rm and}
\;\;\;\;\;\;\; a=\sqrt{\frac{2 \ \tau_{min}}{\alpha}} \ \eta.
\end{equation}
Following the lines developed in Ref.\cite{ls}, a computation of the average
increase of the orbit's action when the volume $V$ of the nucleus changes by
$\delta V$ gives $\eta = \ef \delta V/V$. Computing $\delta V$ when a neutron
is added, and using the expression $\tau_{min}/\alpha = (\pi \sigma_x)^{-2}$
\cite{lm}, where $\sigma_{x} = \langle (\partial \widetilde{U} /\partial x )^2
\rangle^{1/2}_{x}$, we obtain $a = (\sqrt{2}/\pi A) \ \ef /\sigma_{x}$. A
reasonable estimate for $\sigma_x$, that has been tested numerically, is $\sim
3/A^{1/3}$ MeV. Then $a \sim 37/(2 \pi A^{2/3})$, which is of order $0.5$ for
$A\approx 50$ and $0.2$ for $A\approx 200$. We have verified that, for these
values of $a$, the error introduced by setting $a=0$ in Eq.(\ref{cna}) is less
than $7 \times 10^{-2}$ for any $\zeta$. Setting $a$ to zero, the final result
can be expressed as
\begin{equation} \label{cf}
C (\zeta)= \left( 1 - \frac{\zeta^2}{4} \right) \ e^{-\zeta^2 /4} +
\frac{\zeta^4}{16} \ \Gamma (0,\zeta^2 /4) \ ,
\end{equation}
where $\Gamma (s,z) = \int_z^\infty t^{s-1} e^{-t} d t$. Through the
reparametrization (\ref{unf}), all the system specific features have been
incorporated in $\zeta$. This leads to a ``universal'' function,
Eq.(\ref{cf}). Why it is so, as well as the validity of this result, is
discussed below \cite{reg}.

The parameter $\alpha$ contains detailed physical information related to the
single--particle spectrum \cite{ls}. However, it is difficult to extract the
relevant information from experimental data. To circumvent this difficulty,
the alternative expression \cite{lm}
\begin{equation}  \label{unfu}
\zeta=\sqrt{\left\langle \left(\partial_{x_0} \widetilde{U}
\right)^{2}\right\rangle / \left\langle \widetilde{U}^{2}\right\rangle} \ x \ ,
\end{equation}
where $\partial _{x_0} \widetilde{U} = \partial \widetilde{U} (x_0) /\partial
x_0$, is more convenient for our purpose because it only involves properties
of $\ut$. Though the structure of Eq.(\ref{unfu}) is reminiscent of the
reparametrization introduced in the context of random matrices and universal
parametric correlations \cite{sa1,ls}, there are however important
differences:  in the latter case single--particle energies are considered and
universalities are exhibited, whereas here parametric correlations of
thermodynamic properties of a Fermi gas are considered, and important system
specific features are shown to emerge (see below).

From data, we have analyzed the following correlation function
\begin{equation}
C_Z (dN)=\frac{\left\langle \widetilde{U} (Z,N) \ \widetilde{U} (Z,N+dN)\right\rangle
  _{N}}{\left\langle \widetilde{U}^{2}\right\rangle _{N}} \ ,
\end{equation}
where $\widetilde{U} (Z,N)$ is the difference between calculated and measured
masses for a nucleus having $Z$ protons and $N$ neutrons, and $dN$ is the
difference in neutron number along an isotopic chain (a similar analysis for
isotonic chains can also be performed). For a specific $dN$ and $Z$ every
available $f(N)=\widetilde{U} (Z,N) \widetilde{U} (Z,N+dN)$ is calculated. To
compute the mean value with respect to $N$ one has to sum all the $f(N)$'s and
divide by the total number of nuclei in the isotope chain. This normalization
ensures the non--negativity of the Fourier transform of the correlation
function. The average $\left< \widetilde{U}^2 \right>_N$ is obtained by
setting $dN=0$. In order to improve the statistics, that are severely limited
by the length of the isotopic chains, a further average is computed over
neighboring chains,
\begin{equation} \label{cexp}
C(dN)=\left\langle C_Z (dN)\right\rangle _{Z}.
\end{equation}
Finally, to compare with Eq.(\ref{cf}), the correlation $C(dN)$ should be
expressed in terms of the parameter $\zeta$. For each $dN$, (locally) the
value of $\zeta$ is computed from Eq.(\ref{unfu}), where $\partial_{x_0}
\widetilde{U} = \ut (Z,N+1) - \ut (Z,N)$, and averages over $N$ and $Z$ are as
before.

Figure 2 shows the normalized experimental correlation function of the nuclear
mass errors, Eq.(\ref{cexp}), plotted as a function of $\zeta$, compared to
the prediction (\ref{cf}). The three different models studied give very
similar results, in good agreement with theory. The overall quality of the
prediction is comparable to Fig.1. The independence of the result with respect
to the model used is a strong support for our interpretation, though the
influence of effects beyond mean--field theory still need to be clarified.

\begin{figure} \label{corrmassA}
\centerline{\includegraphics[width=7.1cm]{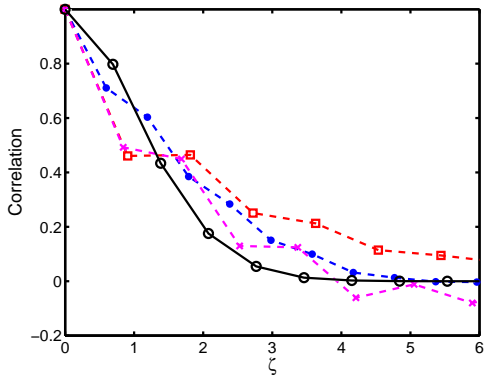}}
\caption{Correlation of the mass errors between neighboring isotopes as a
    function of the dimensionless parameter $\zeta$. Circles, squares and
    crosses correspond to the same models as in Fig.~1. The solid line is the
    theoretical result, Eq.(\ref{cf}).}
\end{figure}

Though Eq.~(\ref{cf}) is a continuous function of $\zeta$ (in particular, it
behaves as $C (\zeta) \approx 1 - \zeta^2 /2$ close to the origin), we have
chosen to plot it at discrete values of $\zeta$, with a step similar to the
experimental one. The half--width of the correlation function $C (\zeta)$ is
of order 1.5, which corresponds to neutron differences $d N \approx 2$, giving
the parameter range over which the chaotic contributions of different isotopes
are correlated. The existence of correlations extending over a few nucleons
seems to be consistent with the high accuracy with which Garvey-Kelson-type
mass relations are fulfilled \cite{bfhv}, as mentioned before.

The main approximation in Eq.(\ref{cf}) is the replacement of the diagonal
form factor, Eq.(\ref{ffd}), by the function (\ref{ffa}). By this, the system
specific spectrum of periodic orbits is replaced by a function that keeps only
one relevant scale, the period $\tau_{min}$ of the shortest one. It is this
simplification that, through the reparametrization (\ref{unf}), allows to
obtain a ``universal'' correlation function $C (\zeta)$. Although this leads
to a reasonable approximation, at least for the available neutron differences,
the exact form of the correlation function of the chaotic contribution to the
mass would be better described by the discrete sum (\ref{cdiag1}). This sum
depends on the precise properties of the periodic orbits, and is therefore
system and model dependent. Because of its discrete nature, generically
oscillations of the correlation as a function of $x$ are expected. This
contrasts with the uniform ``universal'' non--oscillatory decay given by
Eq.(\ref{cf}), which is clearly an artifact of the approximation (\ref{ffa}).
Some tendency towards negative values, and therefore of oscillatory behavior,
seems to be present in the autocorrelation of the errors shown in Fig.2 at
large values of $\zeta$. However, to be conclusive, larger values of $\zeta$,
which are not experimentally available, are needed (this problem might be
overcome by considering different nuclear chains \cite{hfv}).

To conclude, let us recall that shell effects described by periodic orbit
theory related to regular motion are very familiar in nuclear physics,
particularly since the work of Strutinsky (see Ref.\cite{sev} for a recent
discussion). What about the contribution of unstable chaotic orbits? It has
been recently suggested that the size of the differences between measured and
computed binding energies can be attributed to the presence of nuclear chaotic
motion. The work presented here goes one step further in this direction by
studying the autocorrelation function of the chaotic contribution to nuclear
masses. The result, that depends on some well identified physical parameters,
is in good agreement with the autocorrelation computed from several existing
models along isotopic chains. This gives further support to the view that
chaotic dynamics effects are present in the ground state of atomic nuclei.

\noindent $^*$ Unit\'e de recherche associ\'ee au CNRS.


\begin{thebibliography}{99}
\bibitem{reg} The calculation of the autocorrelation for the regular shell
contributions to the mass presents no difficulties since the latter are well
reproduced by a simple model of a spheroidal cavity (see, e.g.
Ref.\cite{sev}).
\end{thebibliography}
\end{document}